\documentclass[draft]{agujournal}
\draftfalse
\journalname{Journal of Geophysical Research: Solid Earth}

\usepackage{graphicx,url,amssymb,amsmath,longtable,rotating,color,units, 
wasysym,subfig,epsfig,multirow}

\begin{document}

\title{Coherence-based approaches for estimating the composition of the seismic wavefield}

%
%

\authors{M. Coughlin\affil{1}, J. Harms\affil{2,3}, D.C. Bowden\affil{4,5}, P. Meyers\affil{6,7}, V.C. Tsai\affil{4}, V. Mandic\affil{6}, G. Pavlis\affil{8},T. Prestegard\affil{6}}
\affiliation{1}{Division of Physics, Math, and Astronomy, California Institute of Technology, Pasadena, CA 91125, USA}
\affiliation{2}{Gran Sasso Science Institute (GSSI), I-67100 L’Aquila, Italy}
\affiliation{3}{INFN, Laboratori Nazionali del Gran Sasso, I-67100 Assergi, Italy}
\affiliation{4}{Seismological Laboratory, California Institute of Technology, Pasadena, CA 91125, USA}
\affiliation{5}{Institute of Geophysics, ETH Zurich, Switzerland}
\affiliation{6}{School of Physics and Astronomy, University of Minnesota, Minneapolis, Minnesota 55455, USA}
\affiliation{7}{Ozgrav, School of Physics, University of Melbourne, Parkville, VIC 3010, Australia}
\affiliation{8}{Department of Geological Sciences, Indiana University, Bloomington, IN 47405, USA}

\correspondingauthor{Michael Coughlin}{mcoughli@caltech.edu}




\begin{keypoints}
\item Correlation measurements indicate that body waves often prevail in ambient seismic noise at 0.2\,Hz
\item Precise prediction of seismic signals using data from an array including underground seismometers is possible
\end{keypoints}

%
%


\begin{abstract}


As new techniques exploiting the Earth's ambient seismic noise field are developed and applied, such as for the observation of temporal changes in seismic velocity structure, it is crucial to quantify the precision with which wave-type measurements can be made. This work uses array data at the Homestake mine in Lead, South Dakota and an array at Sweetwater, Texas to consider two aspects that control this precision: the types of seismic wave contributing to the ambient noise field at microseism frequencies and the effect of array geometry. Both are quantified using measurements of wavefield coherence between stations in combination with Wiener filters. We find a strong seasonal change between body-wave and surface-wave content. Regarding the inclusion of underground stations, we quantify the lower limit to which the ambient noise field can be characterized and reproduced; the applications of the Wiener filters are about 4 times more successful in reproducing ambient noise waveforms when underground stations are included in the array, resulting in predictions of seismic timeseries with less than a 1\% residual, and are ultimately limited by the geometry and aperture of the array, as well as by temporal variations in the seismic field. We discuss the implications of these results for the geophysics community performing ambient seismic noise studies, as well as for the cancellation of seismic Newtonian gravity noise in ground-based, sub-Hz, gravitational-wave detectors.  
 
\end{abstract}

%
%

\section{Introduction}


Significant effort has been made in the wider seismological community to understand and exploit background ambient seismic noise. One important mechanism for the generation of seismic noise relates to continuous harmonic forcing of ocean waves as they interact with both the seafloor and coastlines, and this varies strongly in time, frequency and azimuth \citep[][]{Lon1948}. These mechanisms most strongly generate energy in the range of 0.06-0.13 Hz (8 to 16 second periods), but a much wider range of periods is also observed worldwide \cite[e.g.,][]{Ebl2012}. There can be strong body-wave components as well \cite[e.g.,][]{GeSh2008}. Efforts to image these noise sources usually use array processing methods that consider the coherence of wavefronts incident upon the array, referred to as beamforming or frequency and wavevector ($f$-$k$) analysis \citep[e.g.,][]{Rost2002,GeSh2008}, and share a common goal with the approach outlined in this paper. 

Particular attention has been paid to understanding the effect that the inhomogeneous distribution of noise sources would have on the coherence or cross-correlation measured between stations, with the goal of determining whether measurements can be reliably used for the study of seismic velocities  or attenuation \citep[e.g.,][]{Cupillard2010,Weaver2011a,Tsai2009,Tsai2011,LaPr2011,HaRy2010,YaRi2008}, with additional studies  exploring the extent to which signal preprocessing can reduce the effect of inhomogeneous noise sources \citep[e.g.,][]{BeRi2007,Viens2017}. Some of these velocity or attenuation measurements require a great amount of precision and stability over time \citep[][]{Froment2010}, such as for the observation of material velocity changes; velocity variations on a daily or monthly timescale may be as small as a couple percent, but have been shown to yield valuable information regarding temperature or pore pressure changes \citep[i.e.,][]{Bre2008,Lecocq2014,Taira2016}. This paper explores two aspects of such cross-correlation or coherence based observations that affect the final precision with which measurements may be reliably made. 

The first is an analysis of the types of waves that constitute the background ambient noise field. Should the relative contributions of body-wave energy compared to surface-wave energy change over time, this may bias the velocities measured from coherence or correlation techniques, especially when the inter-station distance is small enough that different seismic phases are not well separated. 
We specifically explore the oceanic microseisms, known to be generated by ocean waves between 50\,mHz--0.3\,Hz.
These include the primary microseismic peak that is commonly accepted to be caused by ocean waves generating pressure in shallow waters near the coast. The primary microseisms define a noise peak at frequencies below 0.1\,Hz. The secondary microseismic peak is commonly thought to be created by two counter-propagating wave fields forming standing waves that define a peak around a 8\,s period. Rayleigh waves are generally observed to dominate the ambient noise field and provide a useful tomographic tool \citep[e.g.,][]{Shapiro2004}. Other authors have noted the presence of P waves \citep[e.g.,][]{Vinnik1973,GeSh2008,LaEA2010,NeHa2018} and S waves \citep[e.g.,][]{NiTa2016} through various cross-correlation or beamforming studies at certain frequencies. Similarly, in this study, coherence measurements are considered in the wavenumber-frequency domain as a function of station-station distance and in the time-domain. We find that for the secondary microseism at 0.2 Hz, differing velocities are observed over the course of a year that can only be explained by differences in the type of wave dominating the measurements. This conclusion that body waves are not only present, but often dominate the wavefield at this frequency, has strong implications for the reliability of coherence-based velocity observations and indicates that care should be taken if measurements are to be made in particular seasons. Blindly averaging noise correlations over the course of a year may give unexpected results under a noise field changing in this manner.

The second analysis considers the geometry of the array being used, and the lower limit to which the wavefield can be adequately resolved. Specifically, we explore the utility of adding underground seismometers as compared to most seismic arrays which are constrained to observations at the Earth's surface. This characterization is done through the construction of ``Wiener filters,'' which simultaneously use coherences between all stations in an array rather than on a station-station basis. Wiener filters are optimal linear filters designed to cancel noise; the extent to which ambient noise can be predicted and subtracted from a given target station directly relates to the array's efficacy at describing the wavefield under changing conditions. This approach is also employed in the marine community with the use of vertical strings of hydrophones \citep[e.g.,][]{Cox1973,VeWi1985,RoHi2012,OzGe2017}, and in the gravitational-wave community where seismic motions need to be subtracted from other measurements \citep{CoHa2018,DaMa2018}. Using underground stations is shown to improve Wiener filter predictions by at most a factor of 4 (the improvement maximal at the microseism), suggesting that the resolution of coherence- or correlation-based imaging can be significantly improved by including underground sensors.  

\begin{figure}[t]
 \includegraphics[width=0.75\textwidth]{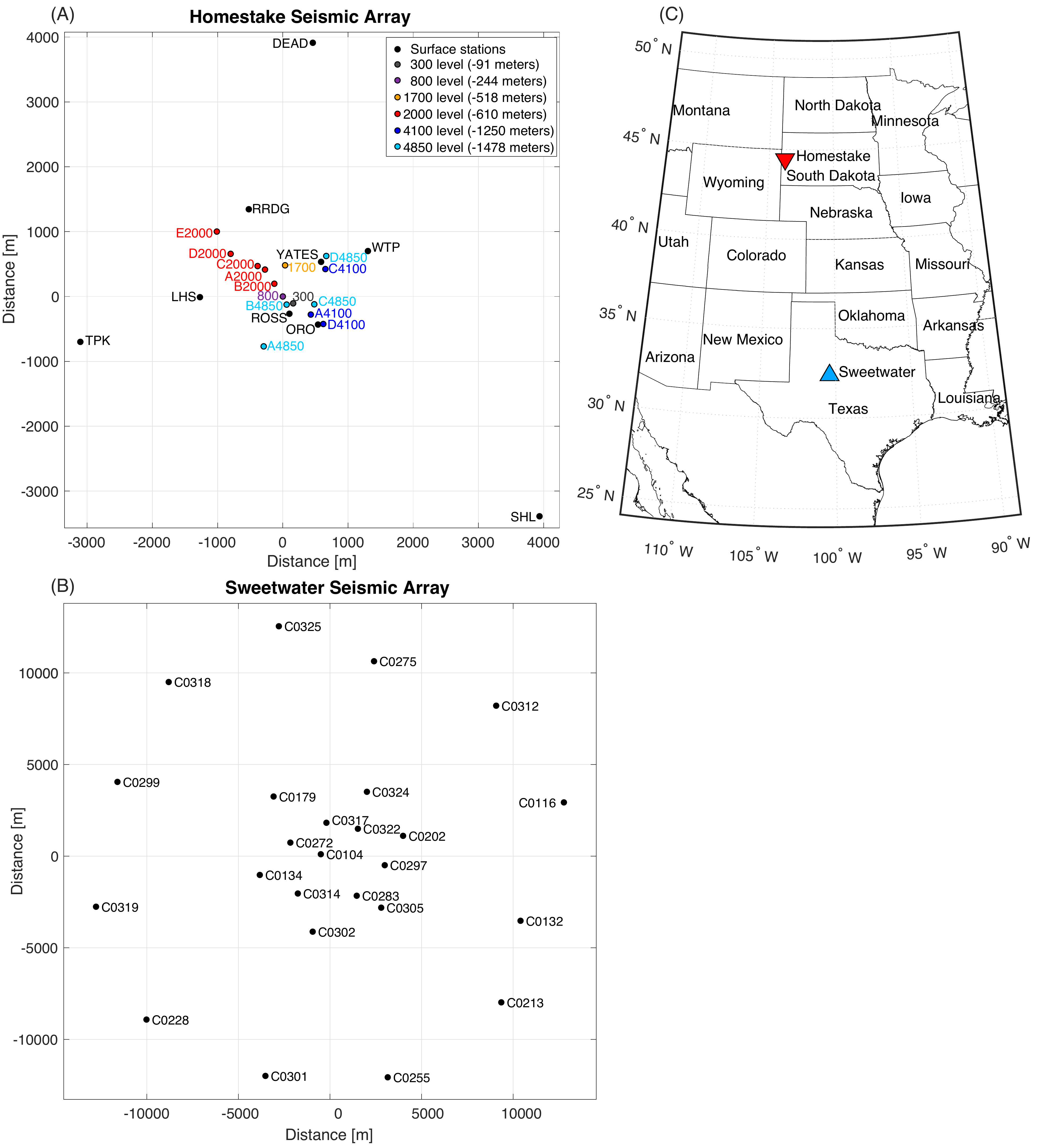}
 \caption{Array geometry and geography, including the names of the seismic stations in the Homestake and Sweetwater networks. For a version of the plot with the local topography and mine drifts, please see Fig.~1 in \citealt{MaTs2017}.}
 \label{fig:array}
\end{figure}

For most of this analysis, we focus on a new seismic array at the former Homestake mine in Lead, South Dakota. Since mining activity has ceased, the Sanford Underground Research Facility there has been demonstrated to be a world-class, low-noise environment \citep{HaEA2010,CoHa2014,MaTs2017}. In 2015 and 2016, a PASSCAL array of 24 broadband instruments (15 underground and 9 above ground) were deployed in and around the mine, covering horizontal distances of more than 6000\,m, and vertical depths of about 1500\,m, shown in Figure~\ref{fig:array}.  The quiet environment and 3D geometry make the array an ideal location to test the approaches and questions described above. We supplement the array data with additional data from a nearby station of the Global Seismographic Network (station RSSD). Finally, an array of instruments in Sweetwater, Texas \citep{BaHo2014} are also briefly used as examples to show that conclusions regarding the wavefield composition are not solely constrained to the Homestake array in South Dakota. The data used span slightly over one year of time, from June 2015 to September 2016.

Finally, we note that the results of this paper have implications for seismic noise and Newtonian gravity-noise reduction in the gravitational-wave community and this is also briefly discussed.

\section{Velocity measurements and wavefield decomposition}
\label{sec:correlation}

This section considers velocity observations made through different approaches.
Observations in this section are made by considering station-station coherence. This is equivalent to the Fourier transform of the stacked cross-correlations used by other studies \citep[e.g.,][]{Aki1957,Boschi2012}, and we define our observations formally here.
The first step is to calculate the complex spectral coherence of all of the vertical channels of seismometer pairs using extended time periods of data. The coherences between seismometers $i,\,j$ were collected over several months in their complex form
\begin{equation}
\gamma_{ij}(f) = \frac{\langle x_i(f)\,x^*_j(f)\rangle}{\sqrt{\langle |x_i(f)|^2\rangle\langle |x_j(f)|^2\rangle\phantom{\big]}}}
\label{eq:coh}
\end{equation}
where $x_i(f)$ is the value of the Fourier Transform at a particular frequency $f$ for the $i$th seismometer, $x^*_i(f)$ its complex conjugate, and $\langle \rangle$ indicate an average over consecutive time windows. This metric keeps information about relative phases between the records of seismometers through phase multiplication. 

Assuming that all seismic sources are sufficiently distant, we can divide the seismic field into four components: shear waves, compressional waves, and surface Rayleigh and Love waves. Our goal is to obtain speed estimates by observing the ambient seismic field. In this case a challenge is that there can be multiple waves contributing simultaneously at all frequencies. The array dimension, i.e., the array size and density of instruments, then sets a lower and an upper limit on the range of frequencies where multiple waves can be disentangled to obtain well-defined differential phases between sensors. 

\subsection{Observations in Frequency-wavenumber Domain} 

Our first estimate of wavespeed is done in the frequency domain using ``$k$-$f$ maps,'' which effectively search for plane-waves of varying direction and speed, testing the total coherence of measurements after the appropriate phase-delays are applied \citep{Rost2002}. Given the distance in 3 dimensions between seismometers, $\vec r_{ij}$, the unshifted station-station coherence $\gamma_{ij}(f)$ (from Eq.~\ref{eq:coh}), and a given wave vector $\vec k(f)$ to test, the probability of a wavefront propagating with that wavenumber is:  

\begin{equation}
m(\vec k,f) = \sum\limits_{i,j}\gamma_{ij}(f)\,\textrm{e}^{\textrm{i}\,\vec k(f)\cdot \vec r_{ij}}.
\label{eq:kfmap}
\end{equation}

\begin{figure}[t]
 \includegraphics[width=0.75\textwidth]{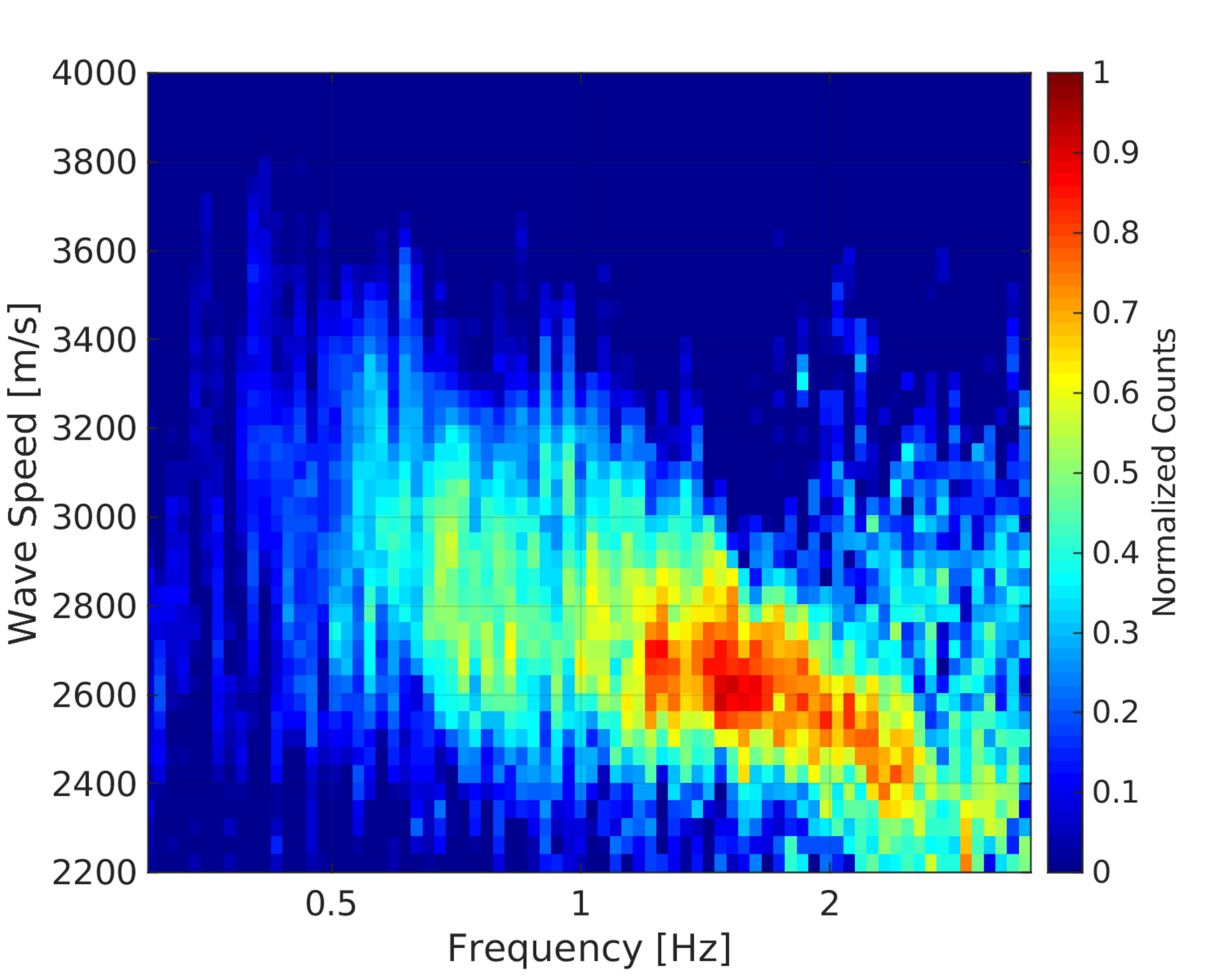}
 \caption{
A histogram of estimated wave speeds between 0.3 -- 3.5\,Hz. Red color means that the respective speed value was measured for a large number of daily $k$-$f$ maps, while blue color means that the speed value was measured rarely.}
 \label{fig:speeds}
\end{figure}

Such array processing approaches have been previously used to explore seismic sources \citep[e.g.,][]{GeSh2008,NeHa2018}. As opposed to analyzing each data stretch individually, we are mostly interested in the background noise field.
For this reason, we average observations over the course of a given day.
To do so, we calculate Eq.~\ref{eq:kfmap} in 128\,s time windows with no overlap.
We note that there is no averaging in Eq.~\ref{eq:coh} in this case.
The values for $m(\vec k,f)$ are averaged over the course of a day.
We sample from $m(\vec k,f)$ to determine $\vec k(f)$, whose values are collected over the course of a year.
We convert wavenumber in $x$, $y$, and $z$ to a total velocity to produce the histogram shown in Figure ~\ref{fig:speeds}. 

\begin{figure}[t]
\includegraphics[width=0.5\textwidth]{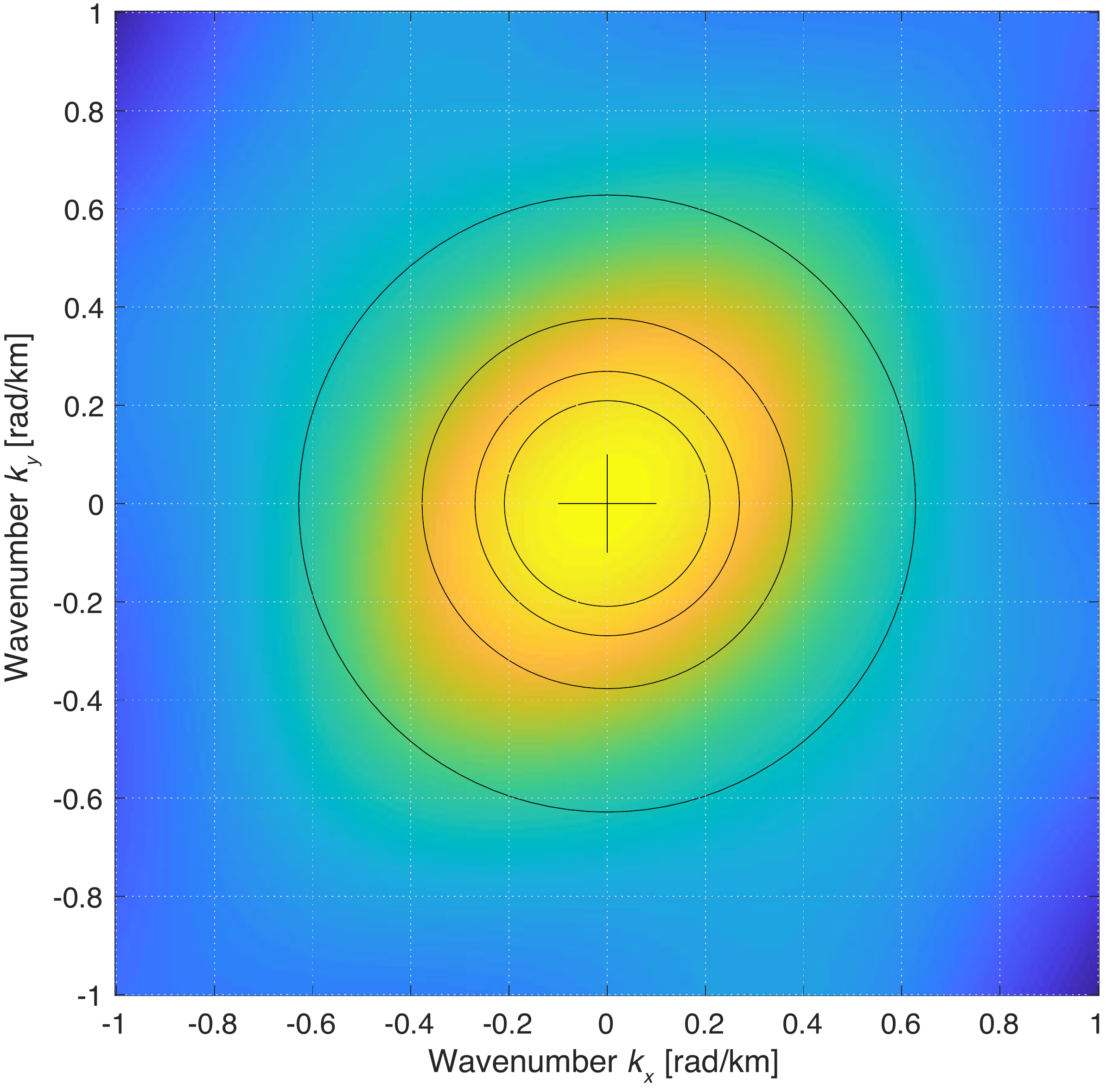}
 \caption{Horizontal wave number response (computed using Eq.~\ref{eq:kfmap} assuming $\gamma_{ij}(f)=1$) of the Homestake array at 0.3 Hz. This only considers $k_z=0$, which is necessary for visualization purposes given the 3D geometry of the array, though we note that we do account for depth in the full estimation of wave speed. The black circles indicate constant speeds of 3, 5, 7, and 9\,km/s.}
 \label{fig:kfmap}
\end{figure}

Figure ~\ref{fig:speeds} shows seismic speeds in the range between 0.3\,Hz to 3.5\,Hz. The distribution of maxima tends to lower speed values at higher frequencies, following the expected dispersion of Rayleigh waves. Between 1\,Hz and 2\,Hz, Rayleigh-wave speed is found to be about 2.6\,km/s falling to lower values above 2\,Hz. At 2.5\,Hz, seismic wavelengths are about 900\,m, which is smaller than the distance between most station pairs. This explains why accurate speed estimates cannot be obtained at higher frequencies. At 0.3\,Hz, wavelengths are about 10\,km, which is longer than the array aperture, and therefore standard speed estimation methods fail at lower frequencies.

We plot the array response function, computed using Eq.~\ref{eq:kfmap} assuming $\gamma_{ij}(f)=1$, at 0.3 Hz in Figure ~\ref{fig:kfmap}, showing that distinguishing wavespeeds at even lower frequencies would be difficult. Therefore while we are confident that the wavefield above 0.3\,Hz is dominated by surface waves, we must turn to alternate methods or use different stations to investigate the wavefield at lower frequencies. 

\begin{figure*}[t]
\centering
\hspace*{-0.5cm}
\includegraphics[width=0.99\textwidth]{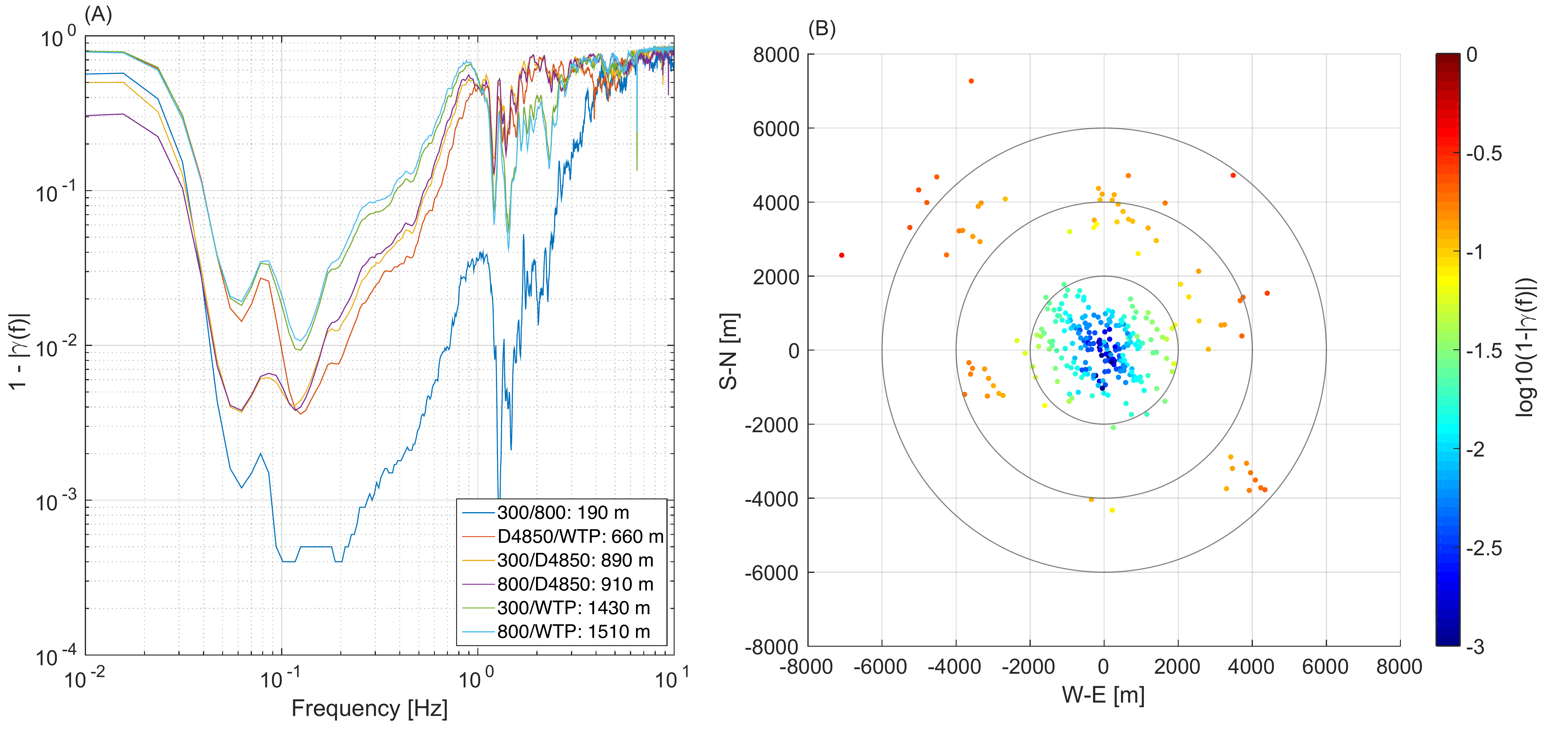}
 \caption{
(A) $1-|\gamma(f)|$ between a variety of seismometer pairs averaged over 6 months of coincident data. The station names with numbers indicate the station depth in feet (we use feet because of long-standing naming conventions in the mine for levels serviced by shaft elevators) \citep{MaTs2017}. The legend indicates the horizontal distance in meters between each pair shown, and the pairs are shown in ascending order of horizontal distance. (B) logarithm of $1-|\gamma(f)|$ at 0.2\,Hz between all seismometers, where the x,y-coordinates correspond to the relative horizontal position vector between two seismometers. We show this version to highlight the values of $|\gamma(f)|$ nearest to 1. 
}
 \label{fig:coh}
\end{figure*} 

\subsection{Coherence Decay with Station-Station Distance}

We can also explore the strength of coherence ($\gamma(f)$ in equation \ref{eq:coh}) as a function of frequency and station-station distance. As we will show, this allows us to characterize the wavefield at lower than 0.3\,Hz despite the relatively small aperture of the array. Here, coherence was calculated with 50\% overlap, and in this form also used later for the Wiener filter section. Coherence is considered for all station-station pairs, and Figure~\ref{fig:coh}A shows the difference $1-|\gamma(f)|$ for a few pairs (shown to highlight the values of $|\gamma(f)|$ nearest to 1). Accordingly, coherence is generally high within the band of the primary and secondary oceanic microseismic peaks between 50\,mHz and 0.3\,Hz, and is insignificant above a few Hertz. Horizontal distances between the seismometer pairs are shown in the legend.
Figure~\ref{fig:coh}B shows the logarithm of $1-|\gamma(f)|$ at 0.2\,Hz for day 191 of year 2015 in a scatter plot where the two coordinates are the components of the relative horizontal position vector between two seismometers. We highlight 0.2\,Hz because it is the most coherent frequency in the array, as can be seen in Figure~\ref{fig:coh}A, and also is the strongest contributor of seismic noise \citep{Pet1993}. We do not include a third coordinate for depth because Rayleigh waves produce displacements whose phase does not depend on depth (although the relative body wave contribution may change with depth). Coherence is well characterized by the horizontal distance between seismometers. There are no major inhomogeneities or outliers from the overall pattern, but close inspection of the plot reveals significant directional dependence approximately aligned with the north-northwest-south-southeast and west-southwest-east-northeast directions.

\begin{figure*}[t]
\includegraphics[width=0.99\textwidth]{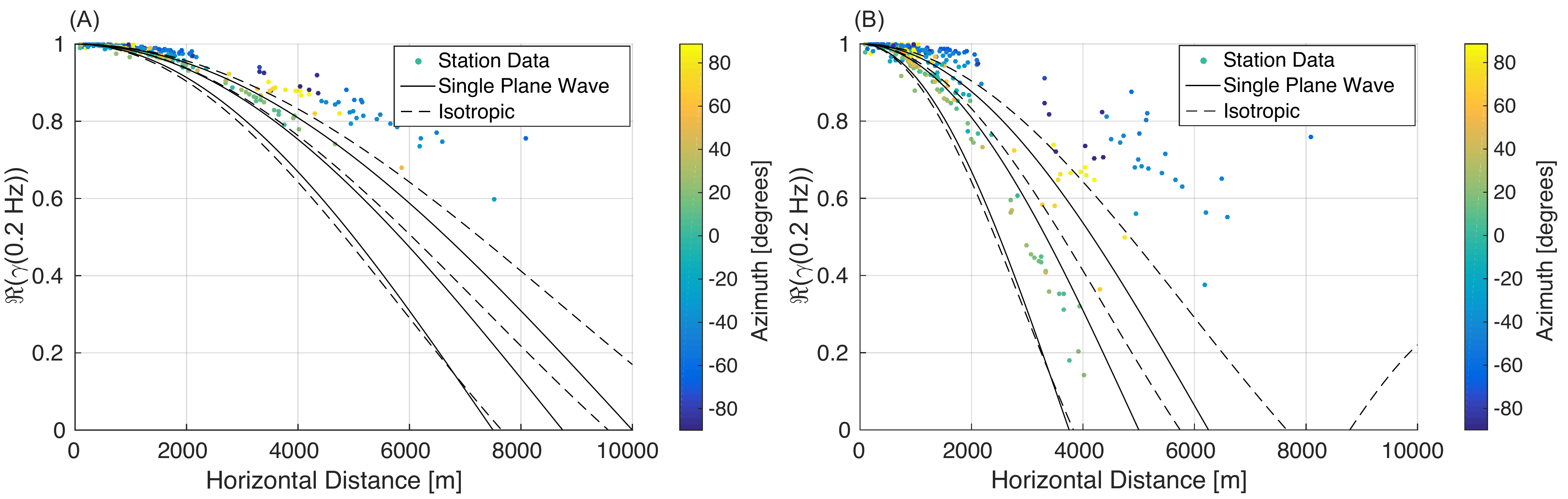}
 \caption{
The RPCC as a function of distance between the vertical channels of all seismometers at 0.2\,Hz for (A) day 154 of 2015; (B) day 191 of 2015. The colors correspond to the azimuth with respect to the east direction of the line connecting two seismometers. In panel A, models are shown for the single plane wave (solid lines; speed values 6\,km/s, 7\,km/s, 8\,km/s), and isotropic field (dashed lines; speed values 4\,km/s, 5\,km/s, 6\,km/s). The same models were used in panel B with speed values 3\,km/s, 4\,km/s, 5\,km/s (solid lines; single plane wave), and 2\,km/s, 3\,km/s, 4\,km/s (dashed lines; isotropic). The velocities are chosen to be consistent with body waves (A) and fundamental Rayleigh waves (B).}
 \label{fig:cohdistance}
\end{figure*}

The rate at which coherence decays as a function of distance can also be used to place constraints on the seismic velocities \citep[e.g.,][]{Aki1957}, and therefore on the composition of the wavefield. We note that \cite{HiRo2016} also provides a similar implementation of this idea to measure a wavefield and the spatial extent to which it collapses to the ideal zero-lag coherence decay. In our case, we note that the decay rate depends on assumptions about the background ambient noise field, so we specifically focus two end-member, orthogonal models for the wave field (there is a third discussed in \cite{Cox1973}, which is a 3-D distribution of plane waves). The first model assumes that the noise field is composed of plane waves that are uniformly distributed in azimuth for a given phase velocity $c$, which would imply that the real part $\Re(\gamma)$ of the complex coherence (RPCC) is given by $J_0(2\pi r/\lambda)$, where $J_0$ is the Bessel function of order zero and $\lambda$ is the wavelength of the waves \citep{Aki1957}. The second end member is the possibility that the wave field is composed of a single plane wave where an angle $\theta$ is the azimuth of the source relative to the station pair. This results in a RPCC of $\cos(2 \pi \cos(\theta) r/\lambda)$. We can take the point at which $\Re(\gamma)=0.5$ as a diagnostic point for this function. For an isotropic Rayleigh-wave field, this value is observed at a distance $r = \lambda/4$. On the other hand, for the plane wave case, the distance between the seismometers needs to be at least as large as $r=\lambda/6$ to observe $\Re(\gamma) = 0.5$. Equality is reached for $\theta=0$\,degrees, and distances of $r>\lambda/6$ are possible in the case of seismometer pairs separated along different directions. 

We plot the RPCC in Figure \ref{fig:cohdistance} at 0.2\,Hz for the two days 154 (A) and 191 (B) of year 2015. These days are during the summer time, but we have checked that similar patterns are also observed in the winter time. The plots show a bimodal distribution, which is a consequence of the directional dependence of the seismic field together with the directional non-uniformity of the seismic array. A uniform array would lead to a continuous distribution of RPCC values. The directional dependence of the seismic field is expected from the known distribution of sources of oceanic microseisms \citep[e.g.,][]{Stehly2006,Harmon2010}, and previously observed at Homestake \citep{HaEA2010}. Extending the lower envelope of the scattered points in Figure~\ref{fig:cohdistance}A with an isotropic correlation model to a coherence value $\Re(\gamma)=0.5$, we find for day 154 that the minimal distance with $\Re(\gamma)=0.5$ is about 7\,km. Using both an isotropic and single plane wave model, we find $\Re(\gamma)=0.5$ is about 3\,km for day 191. Assuming isotropy, we can infer for day 154 a seismic speed of about $\rm 4\times 0.2\,Hz\times 7\,km = 5.6\,$km/s. A similar calculation gives 8.4\,km/s assuming maximal directional dependence. The corresponding values for day 191 are 2.4\,km/s and 3.6\,km/s. While the speed values of day 191 are consistent with expected fundamental Rayleigh-wave speeds, the inferred speeds of day 154 are too high. This implies that non-horizontally traveling body waves must dominate the ambient noise field observed on day 154. 

The bimodal distribution of coherence values in Figure \ref{fig:cohdistance}B is explained by a combination of a non-uniform distribution of wave speeds in the seismic field and non-uniformity of the array. Almost all of the pairs in Figure \ref{fig:cohdistance} with horizontal distance $>2\,$km include a surface station since surface stations are generally located at a greater distance from the main underground array. Surface stations TPK, WTP, and LHS lie on a line pointing approximately along the E-W direction, while the line DEAD-SHL is almost perpendicular to it. Identifying seismometer pairs of the $>2\,$km coherence values, we find that SHL and DEAD appear in the high-coherence part while TPK, LHS, and WTP appear in the low-coherence part. This is consistent with a directional dependence of a seismic field consisting mainly of waves propagating along the E-W direction (roughly towards the Pacific and Atlantic oceans), and the bimodal structure is enhanced by the approximate cross-shape of the surface array. 


We can also exclude any significant impact from transient local sources at 0.2\,Hz irrespective of whether they produce coherent or incoherent disturbances between stations. Observations covering the western US showed that speeds of fundamental Rayleigh waves with an 8\,s period are about 3.1\,km/s \citep{Shen2013}. Together with our results in Figure \ref{fig:speeds}, we can infer that Rayleigh-wave speed at 0.2\,Hz should have wavelengths larger than the array dimension. We also checked that coherence does not decrease systematically when increasing correlation time from one day to one month or longer. This means that there are no significant incoherent disturbances that would average out over long periods of time. Also, we know from our observation of seismic spectra that local disturbances must be weaker than oceanic microseisms by a factor 10 or more as we can see no disturbance visible in time-frequency spectrograms even when oceanic microseisms are close to their minimum.
These observations of coherences and seismic velocities imply that during day 154, the dominant contribution to the seismic field comes from body waves, while Rayleigh waves dominate on day 191. 

\begin{figure*}[t]
  \includegraphics[width=0.9\textwidth]{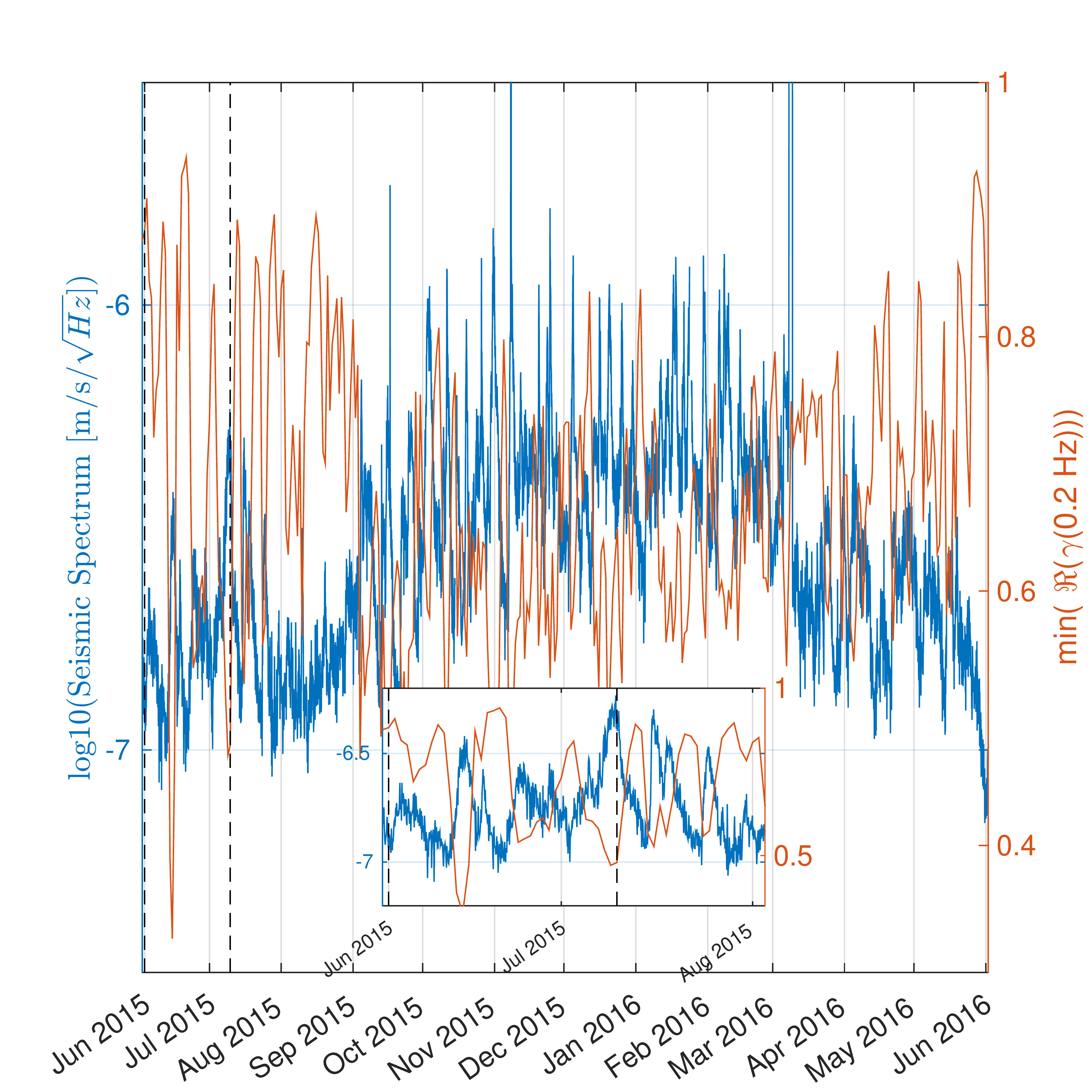}  
 \caption{
The plot shows the power spectral density (PSD) of the 800\,ft station in the vertical direction at 0.2\,Hz and the minimum coherence among all station pairs whose horizontal distance is less than 3\,km, where the dashed vertical lines mark the days 154 and 191 of year 2015 which are used in coherence plots shown in Figure~\ref{fig:cohdistance}.}
 \label{fig:cohpsd}
\end{figure*}

\begin{figure*}[t]
  \includegraphics[width=0.9\textwidth]{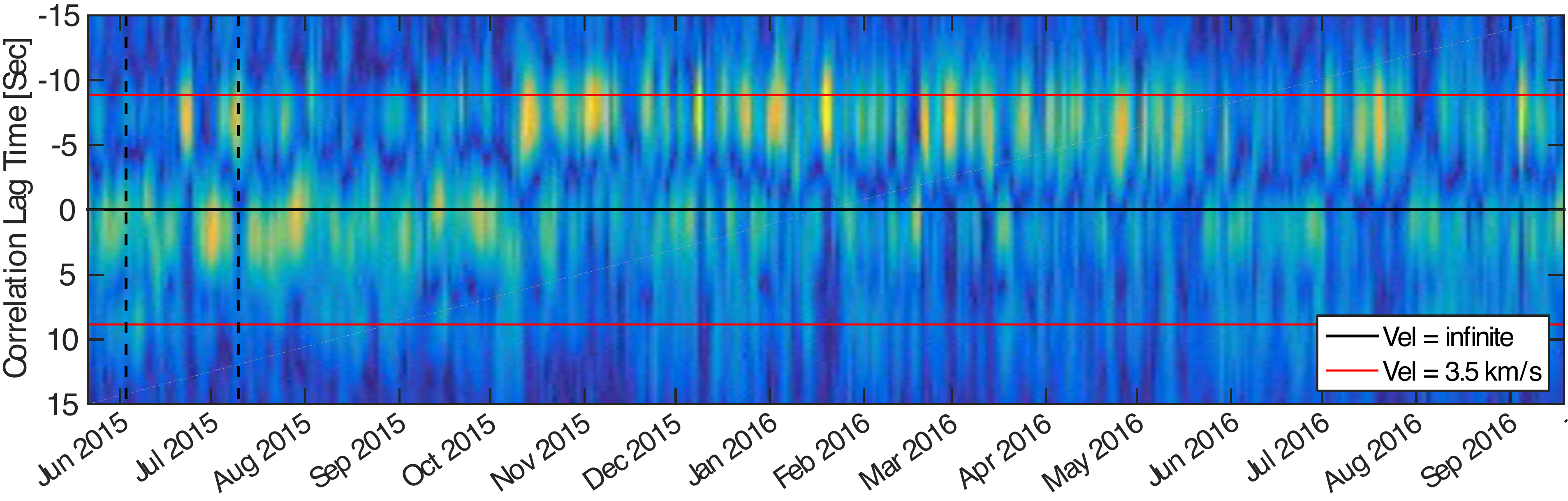}  
 \caption{
The plot shows the bandpassed noise-correlation function between one of the Homestake surface seismometers (SHL) and a nearby instrument from the Global Seismograph Network (RSSD) roughly 31 km away.}
 \label{fig:envelope}
\end{figure*}

To consider a wider range of times, Figure \ref{fig:cohpsd} shows the PSD of the 800\,ft station at 0.2\,Hz over one-year together with the minimal coherence observed between all seismometer pairs closer than 3\,km to each other. The inset plot zooms onto the first 60 days. The expected coherence from an isotropic fundamental Rayleigh-wave field with a speed of 3.5\,km/s (among all plane-wave models, the isotropic model has the highest minimal coherence value) between two seismometers at 3\,km distance to each other is 0.73 (assuming negligible instrumental noise). Coherence exceeds this value significantly during many days, and interestingly, a significant decline of coherence is always accompanied with a significant increase of the microseismic amplitude. This anti-correlation provides further evidence that near vertically-incident body waves not only dominate the wavefield, but that they tend to dominate during the days with lowest ambient noise energy. We will further test this hypothesis in the next subsection.

A possible interpretation of these results is that an incessant background of body waves exists with a spectrum close to the global low-noise model occasionally disturbed by stronger Rayleigh-wave transients. The body waves can be produced at great distance to Homestake since they experience weak damping. Therefore, it is conceivable that body waves arriving at Homestake typically originate from a large number of individual sources, which causes the incessant body-wave background. Instead, the Rayleigh-wave transients are typically produced by relatively close ocean wavefields. Rayleigh waves from more distant sources experience too strong damping to contribute significantly to the field at Homestake.

\begin{figure*}[t]
\includegraphics[width=0.99\textwidth]{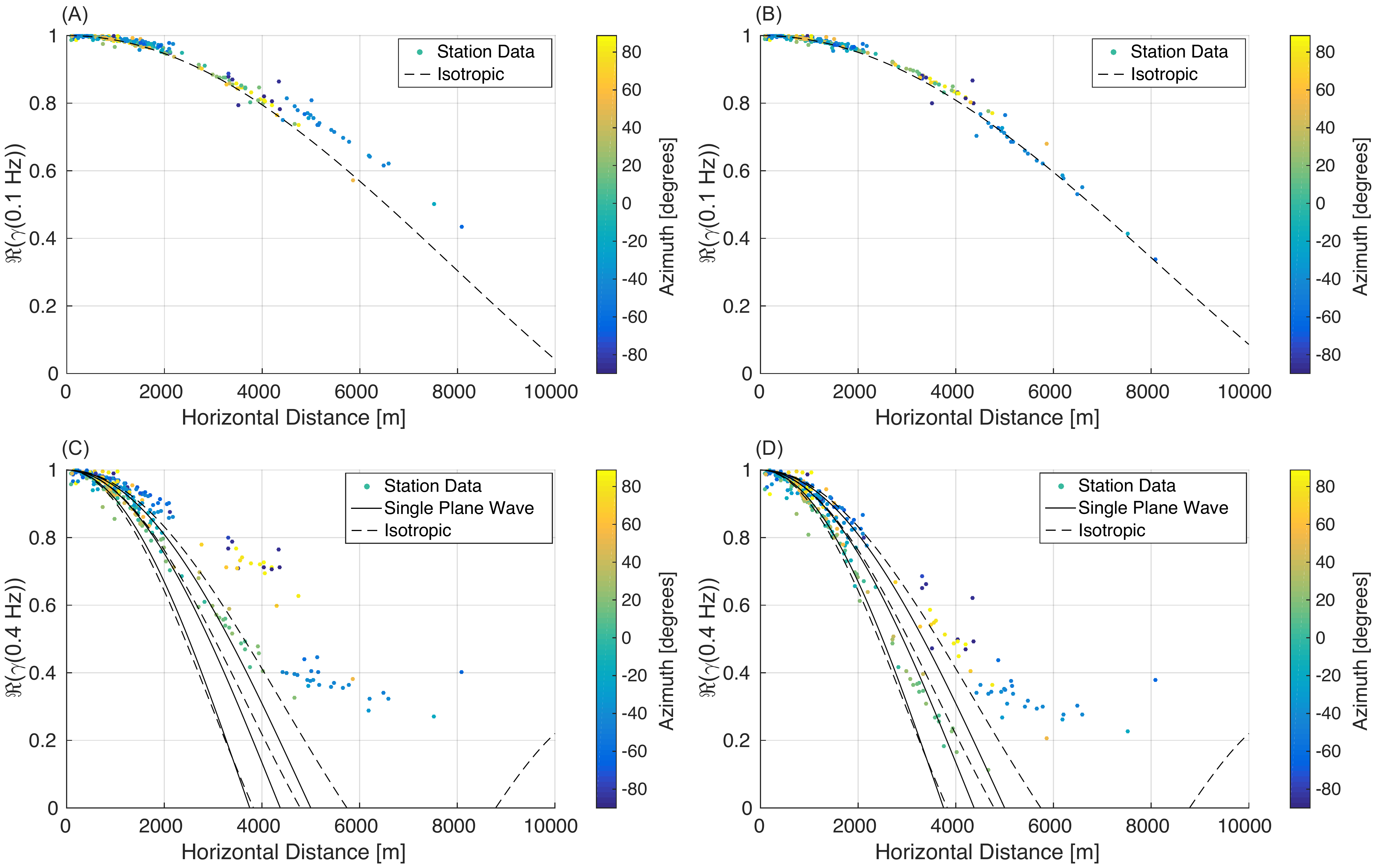}
 \caption{
The RPCC as a function of distance between the vertical channels of all seismometers at 0.1\,Hz (A and B) and 0.4\,Hz (C and D) analogous to Figure \ref{fig:cohdistance}. We use the same days as in Figure~\ref{fig:cohdistance}. In the top row (A and B), an isotropic correlation model is shown with speed value 2.7\,km/s (A), and 2.8\,km/s (B), and in the bottom row (C and D), models are shown for the single plane wave (solid lines; speed values 6\,km/s, 7\,km/s, 8\,km/s), and isotropic field (dashed lines; speed values 4\,km/s, 5\,km/s, 6\,km/s).}
 \label{fig:cohdistancefreqs}
\end{figure*}

\subsection{Time-domain observations}

To further test this observation, we show an alternate version of this analysis.
Figure~\ref{fig:envelope} shows the envelope of time domain cross-correlations between one of the Homestake seismometers (station SHL) and a nearby instrument from the Global Seismograph Network (station RSSD) roughly 31 km away. The correlation for a given day was constructed by averaging hourly coherence measurements between the two vertical channels. This includes a time-domain running-mean normalization and a frequency-domain spectral whitening; these techniques are common in the  community to reduce the influence of earthquakes or other spurious noise sources ~\citep[i.e.,][]{BeRi2007}. The resulting correlation functions are bandpassed from 0.1 to 0.3 Hz. Both positive and negative lag times are plotted, corresponding to coherent signals traveling from RRSD to TPK or from TPK to RSSD, respectively. Horizontal red lines indicate the expected group arrival of surface waves (at either positive or negative correlation lag times) traveling at 3.5\,km/s. While surface waves dominate in the winter months when the 0.2\,Hz microseism noise is strong, many times of the year are dominated by the peak near zero-lag.  Since zero lag implies infinite velocity, this peak is most consistent with body wave arrivals with a high apparent velocity. This, together with the anti-correlation observed between coherence and power spectral density shown in Figure \ref{fig:cohpsd}, suggests body waves incident from below the two stations.

\subsection{Comparisons and discussion of wave content}

\begin{figure*}[t]
  \includegraphics[width=0.9\textwidth]{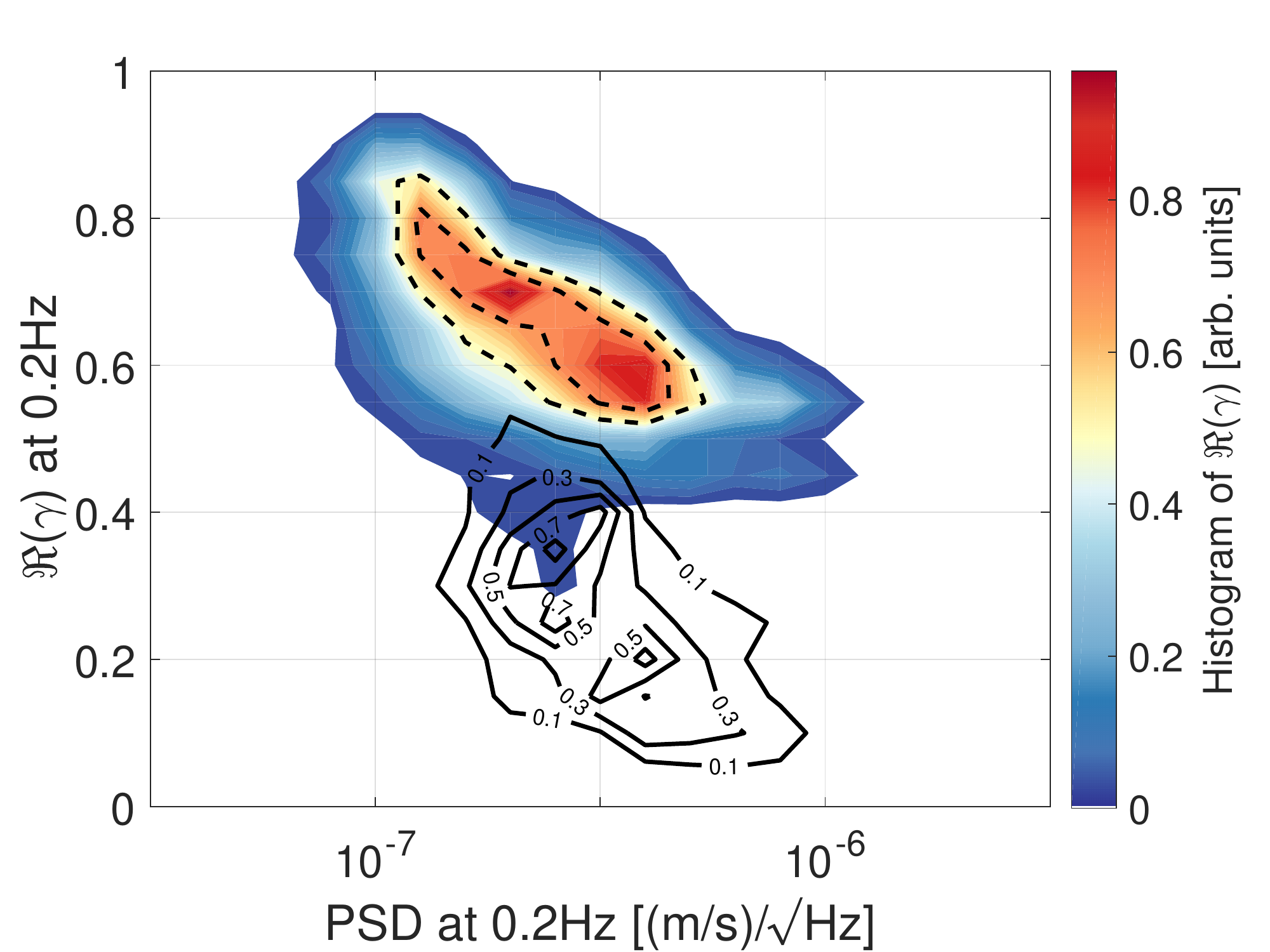}
 \caption{
The plot shows the data in Figure \ref{fig:cohpsd} as a density plot for the Homestake array (colored contours with contour lines at 0.5 and 0.7). The Sweetwater array results are shown with only black contour lines. We use more contour levels in this case to include information lost by not including the colors.}
\label{fig:densitycoh}
\end{figure*}


Our results point strongly towards the following model of oceanic microseisms at Homestake at 0.2\,Hz, and more generally at quiet seismic stations in interior continental regions. When the oceanic microseisms are weak, they approach the global low-noise model \citep{Pet1993}. In this case, the field is dominated by body waves. Typically week-long, strong transients of Rayleigh waves (e.g., from strong Pacific or Atlantic storms) add to this background of body waves, decreasing RPCC values because of the slower velocities of fundamental Rayleigh waves. 
The existence of body waves in oceanic microseisms is well known and modeled previously \citep{GeSh2008,LaEA2010,ObEA2013,NiTa2016}. However, the hypothesis that body waves can dominate the microseismic spectrum at quiet times has not been formulated before to our knowledge. This link seems to exist at the Homestake site at least, and it would be very interesting to obtain direct confirmation using other methods \citep[e.g.,][]{LaEA2010}.
Our method shows that it is possible to differentiate between fundamental Rayleigh waves and body wave contributions; this is potentially important for the field of time-dependent velocity measurement, as well as for ambient noise correlation studies more generally.


The results thus far have focused on observations at 0.2\,Hz, specifically for the Homestake Array in South Dakota. The results, however, can be potentially generalized to other frequencies and locations. To consider other frequencies, we also plot the RPCC at 0.1\,Hz and 0.4\,Hz.
In the top row of Figure~\ref{fig:cohdistancefreqs}~(A and B), we show the RPCC as a function of distance for 0.1\,Hz, and in the bottom row (C and D) for 0.4\,Hz.
We can use the RPCC measurements to constrain seismic velocities at these frequencies as well.
The seismic speeds, measured to be $\approx$\,3\,km/s, are consistent with fundamental Rayleigh waves at 0.1\,Hz.
There is no visible evolution between the days that were dominated by body waves and fundamental Rayleigh waves as in the case of 0.2\,Hz.
On the other hand, results for 0.4\,Hz, at the high frequency end of the microseism, are significantly more complicated. 
Measurements have contributions from both fundamental Rayleigh waves and body waves, and the trend is similar to that of 0.2\,Hz waves. A more systematic analysis over a larger range of frequencies should be a focus of future work (see \cite{ArHe2013} and \cite{TrGe2013} for treatments from a theoretical perspective).

While a study of global patterns is beyond the scope of this study, we can at least examine one other array in the crustal interior of the U.S.
To check that the anti-correlation between PSDs and minimal coherence at 0.2\,Hz is not only present at Homestake, we performed the same analysis for the Sweetwater broadband array \citep{BaHo2014}. The seismometers in this analysis are from an array in Sweetwater, Texas, which is located at 32$^\circ$28'5''\,N and 100$^\circ$24'26''\,W. The array consists of two approximate circles, one with about a 10\,km diameter and another with a 25\,km diameter. We found 23 stations with good data quality during March and April 2014. This array has significantly larger horizontal spacing than the Homestake array, with horizontal distances between the center of the array and other seismometers ranging between 2-14\,km. It also has significant variation in elevation over the array, with a maximum elevation change between seismometers of about 250\,m. We perform the same analysis with this array as in the Homestake case, computing the PSDs and coherences between the station pairs. Figure~\ref{fig:densitycoh} shows $\Re(\gamma)$ vs.\,the PSDs for the Homestake and Sweetwater arrays. Due to the $\approx 10$\,km array extent of Sweetwater as compared with the $\approx 3$\,km extent of Homestake, $\Re(\gamma)$ is expected to be 0.34 for uniformly distributed surface waves instead of 0.73 found at Homestake. Average values of RPCC above this value again suggest body waves are dominant, and the anti-correlation of RPCC with PSD amplitude again suggests that body waves dominate the observed microseism during the lowest amplitude microseism. 

To summarize, we have used coherence measurements coupled with models for the seismic wavefield to constrain the wave types as a function of time at both Homestake and Sweetwater. We have shown that body waves and surface waves contribute energy at different amounts over the course of a year. At 0.2\,Hz, observed velocities shift substantially depending on the season, indicating a dominance of either surface waves or body waves. This may be misinterpreted as a time-varying velocity change for small aperture arrays where seismic phases are not well separated, and implies care should be taken if observations are to be stacked over an entire year or if short deployments are to be used in particular seasons. Many studies assume that Rayleigh waves are responsible for the travel times observed \citep[e.g.,][]{Shapiro2004}, but the mixing of wavetypes may lead to incorrect velocity inferences.

\section{Testing recoverability of the wavefield with Wiener Filters}
\label{sec:wiener}

Given that the ambient noise field is constantly changing in direction and wavetype, there should be a limit in a given array's ability to describe the ambient noise wavefield. That is, should the ambient noise correlation functions from one time period be compared to that of another, there may exist some portion of the wavefield that must be attributed to random, variable processes that cannot be resolved given the geometry of instruments used. This section explores this limit for different array geometries by the construction of Wiener filters. The Wiener filter approach is in many ways similar to the work presented above, but rather than consider only two stations at a time, the Wiener filter simultaneously considers all available station-station coherences from the array. A "target" station is defined, for which information from all others in the array are used to predict and subtract known signals. The extent to which signal remains after this subtraction at subsequent observation times indicates the lower limit to which coherence-based approaches can be reliably interpreted.

In previous work \citep{CoHa2014}, they implemented feed-forward noise cancellation using an array of 3 seismometers in the same general location as our current Homestake array \citep{HaEA2010}. They used Wiener filters, which are optimal linear filters, to cancel noise of (wide-sense) stationary random processes defined in terms of correlations between witness and target sensors \citep{Vas2001}. They explored how to maximize subtraction, including exploring the rate at which the filters are updated and the number of filter coefficients. There were limits to this original study. Due to the fact that they only had three functional seismometers, they could not explore the effect of body waves on the coherence between the seismometers and thus the study of its effects on the subtraction that they could achieve was limited. In addition to the self-noise of the seismometers, topographic scattering and body waves in the seismic field could limit performance \citep{CoHa2012}.

The method is common in gravitational-wave studies, for which the interest is to use arrays of seismometers as witness sensors to the gravitational-wave interferometer to subtract the noise in the seismic field present in the detector. 
The goal of Wiener filtering in this context is to make predictions of the time-series at a single sensor (target sensor) based on observations of other sensors (witness sensors). 
Wiener filtering uses the correlation of all sensors of the array, including accounting for both correlations amongst the witness sensors and the target sensor, when making the predictions. We note that there are other applications that may fall under the classification of a Wiener filter, such as recent work by \citet{Moreau2017} which use similarities in noise correlation functions on different days to successfully extract salient features and de-noise the final stacked observation.
That approach and ours are mathematically similar, but the goal in our case is to reconstruct the raw waveform at a target sensor. In doing this, our filter should encode information about the propagation delay times, amplitude effects, and any other effects from the intervening geological structure.

The method for computing the Wiener filters is as follows.
For samples $y(t_i)$ from a single target channel, $M$ input time series $\vec x(t_i)=(x_m(t_i))$ with $m=1,\ldots,M$, and a Wiener filter $\vec h(i)=(h_m(i)), i=0,\ldots,N$ that minimizes the residual error, the residual seismic time-series can be written symbolically as a convolution (symbol $*$) \citep{Vas2001}:
\begin{equation}
r(t_i) = y (t_i)-\sum\limits_{m=1}^M (h_m*x_m)(t_i),
\label{eq:residualnoise}
\end{equation}
where the convolution is defined as
\begin{equation}
h_m*x_m(t_i) = \sum\limits_{k=0}^N h_m(k) x_m(t_{i-k}),
\label{eq:FIR}
\end{equation}
where $N$ is the order of the finite impulse-response filter $h$ (see section 4.3 of \citep{Orf2007}). This filter depends on the correlations between $y$ and channels $\vec x$ as well as on correlations among channels $\vec x$, but once calculated, it is applied to each channel in $\vec x$ separately as shown in equation (\ref{eq:residualnoise}). In this analysis, we only use past data to construct the current sample.

The resulting set of impulse-responses may be thought of as capturing propagation phase delays, amplitude changes, additional phases from reflections, etc., and the linear combination of each input timeseries convolved with its appropriate filter constructs the target observation as well as possible. In some ways this is similar to beamforming techniques \citep{Rost2002}, which test various incident slownesses by prescribing phase delays between each station, ultimately summing the observations in a linear combination (or in some cases, summing the coherence of each). In the case of beamforming however, only phase delays are considered and not amplitude modulations, and a constant, homogeneous velocity structure is assumed. The Wiener filter approach is agnostic to these assumptions. Additionally, although this study only uses vertical-component traces, multiple seismometer components or even other types of instruments could be included.

It is useful to compare the measured residuals to expected estimates. The extent to which a prediction at a target sensor can be made depends on the station-station coherence observed. These expected residuals can be computed as follows. If we denote $C_{\rm SS}$ as the matrix containing the cross-spectral densities of witness seismometers, $\vec C_{\rm ST}$ as the vector containing the cross spectral densities between the witness and target sensors, and $\vec C_{\rm TT}$ as the PSD of the target seismometer, then the average relative noise residual $R$ achieved is given by
\begin{equation}
R(f) = 1 - \frac{\vec{C}_{\rm ST}^*(f) \cdot C_{\rm SS}^{-1}(f) \cdot \vec{C}_{\rm ST}(f)}{C_{\rm TT}(f)}.
\label{eq:R}
\end{equation}
where superscript * refers to a Hermitian transpose. When using just a single witness seismometer, this simply reduces to
\begin{equation}
R(f)=1-|\gamma(f)|^2
\end{equation}
where $\gamma(f)$ is the witness-target coherence as defined in equation (\ref{eq:coh}). To again compare the Wiener filter approach to beamforming, we note that the construction of a matrix containing all stations' cross spectral-densities, like in $C_{\rm SS}$, is used in both methods. Additional beamforming resolution or other characterizations of the wavefield are possible with various techniques, such as is described by \citet{Cap1969}, MUSIC \citep[e.g.,][]{Goldstein1987,Meng2011}, or an eigenvalue decomposition of the matrix $C_{\rm SS}$ as is explored by \citet{Seydoux2017}, though we do not explore these methods further here. Also, we note that some high-resolution beamforming methods focus on resolving independent transients rather than characterizing the consistent background features as we are interested in here.

\begin{figure*}[htp!]
 \centering
\includegraphics[width=0.80\textwidth]{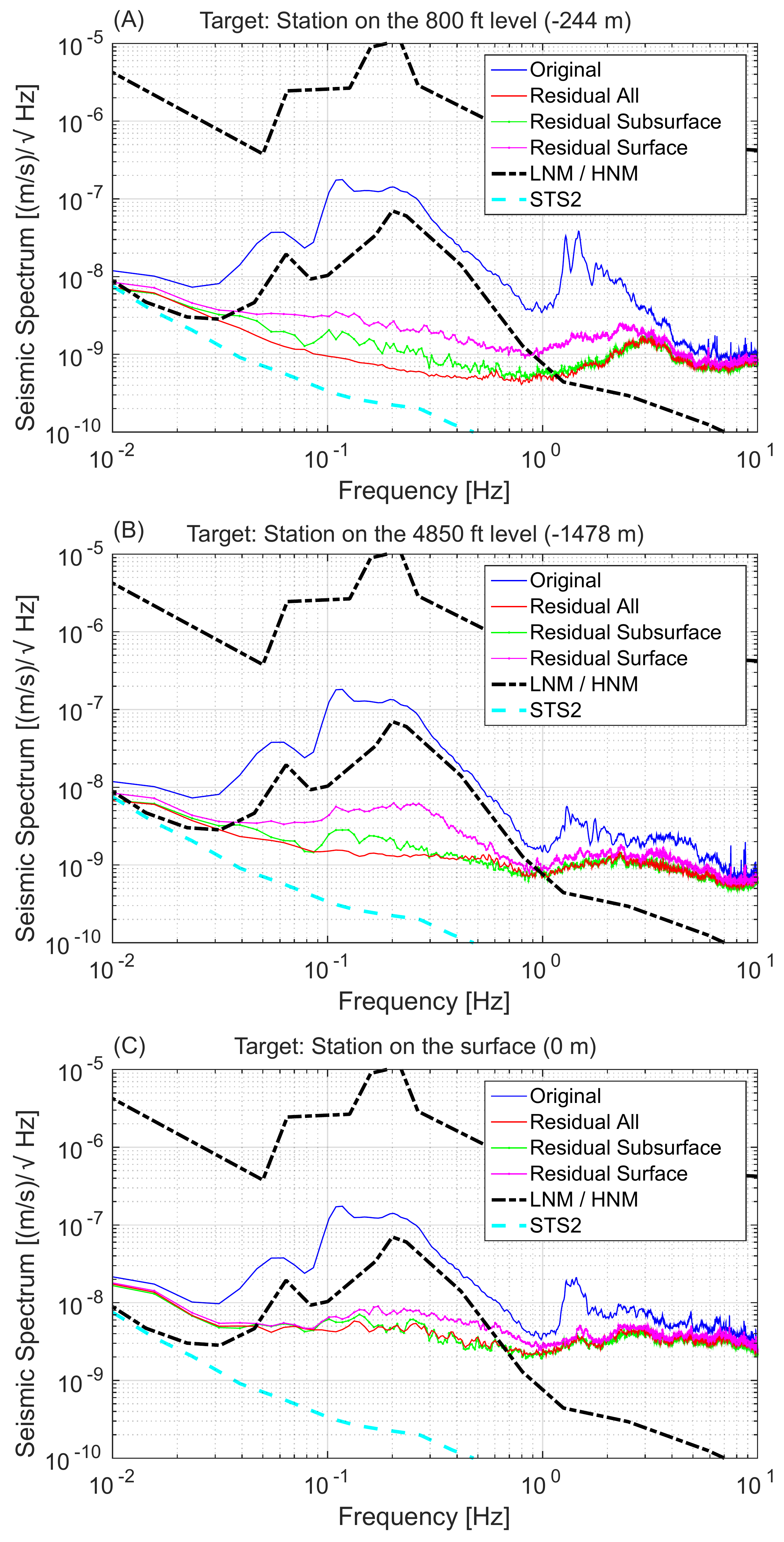}
 \caption{(A) The subtraction achieved using the vertical channel of the seismometer on the 800\,ft level (243\,m depth) as the target channel, up to 10\,Hz. (B) and (C) are the same for the seismometers on the 4850\,ft level (1478\,m depth) and the surface, respectively. In each plot, it is shown how the subtraction varies depending on what set of seismometers are used as witness sensors (subsurface, surface, and all). The dashed black lines correspond to Peterson's high and low noise models \citep{Pet1993}. The instrument noise level for the STS-2 sensors is shown as a dashed green line to show the residual theoretical floor. 
}
 \label{fig:wiener}
\end{figure*}
\clearpage

Since the Wiener filter method makes no assumptions about the wave modes present in the noise field it is important to explore how well it can predict the overall wavefield. If the method is effective, it could open up a large range of future applications of seismic arrays. In the following analyses, we choose one station as our target. The Wiener filter is constructed to predict signals at this station using different subsets of other stations, and then the filter's prediction at subsequent times is compared to actual observation.
Here we show results from analysis of a 3-hour time window without any clear transient signals.  We assume this example is representative.  The broader concept for future applications would be to periodically retrain the filter.  
In Figure \ref{fig:wiener}, we demonstrate the performance of the filter on the seismic array data for different choices of the target station. We achieve more than a factor of 100 reduction in noise at the microseism peak when using all available channels for Figures~\ref{fig:wiener}A and~\ref{fig:wiener}B. Achieving more than a factor of 100 reduction of noise means that we can predict the seismic time-series of the target to better than 1\%.
We can also explore the loss in information from using only surface stations when measuring the seismic wave-field below ground.
We see that the subtraction performance using only surface stations as witness channels is a factor of $\approx 4$ worse than the configuration where all channels are used.
In this way, sub-1\% prediction of the underground seismic wavefield is not possible with only surface sensors.

\begin{figure*}[t]
\centering
\hspace*{-0.5cm}
\includegraphics[width=0.99\textwidth]{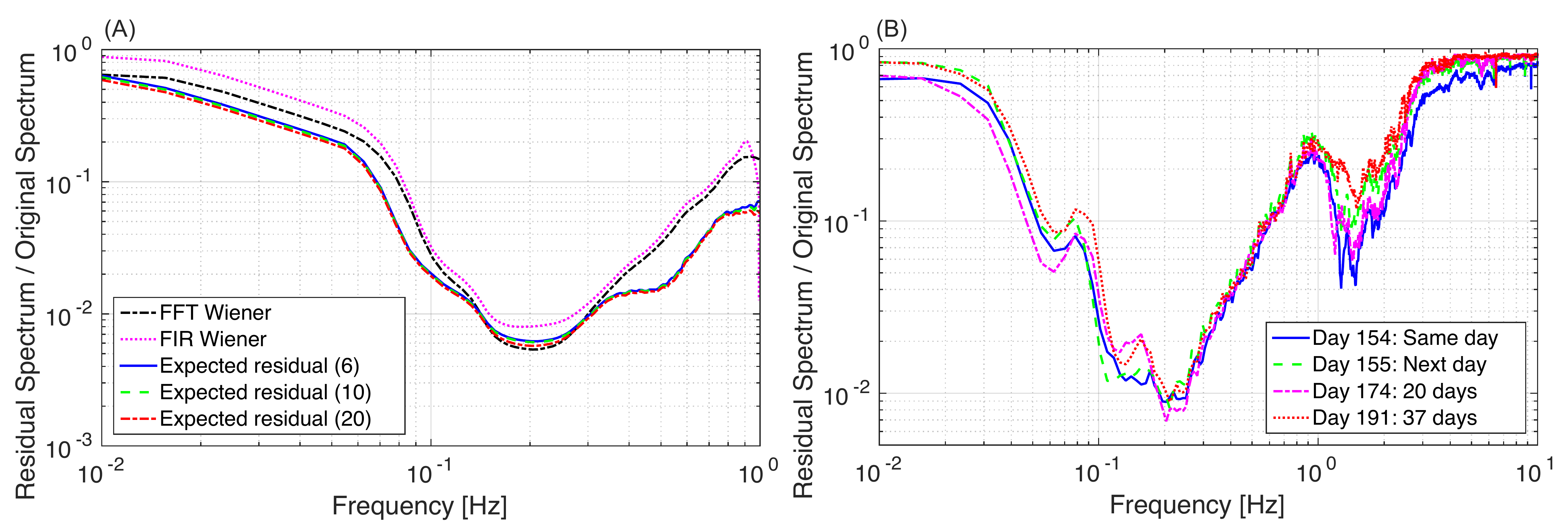}
 \caption{
(A) The expected residuals given the expression in equation (\ref{eq:R}) for a number of seismometer array subsets (the number specified in parentheses in the legend) and comparisons to both FFT Wiener and FIR Wiener filters for the vertical channel of the 800\,ft station as the target channel.
(B) We show the performance of the Wiener filter over a few timescales using the vertical channel of the 800\,ft station seismometer as the target. This result shows that Wiener filters are efficient in this band over long timescales.
}
 \label{fig:longterm_FIRFFTvsFIR}
\end{figure*}

Noise residuals were computed for two different implementations of Wiener filters described earlier. One is the frequency-domain filter \citep{AHO1999}. The other is the finite-impulse response (FIR) filter applied as shown in equations (\ref{eq:residualnoise}) and (\ref{eq:FIR}) of order $N=8$.
Figure~\ref{fig:longterm_FIRFFTvsFIR}A shows the ratio of the original PSD to the PSD calculated from the FFT Wiener and FIR Wiener filters applied to the vertical channel of the 800\,ft station as the target channel. 
The frequency-domain filter typically achieves slightly better cancellation performance than the FIR filter.
The FIR filter, which is applied in the time domain, has to cope with strong correlations potentially between all samples of the time series. This makes it numerically more challenging to calculate the Wiener filter mostly due to large, degenerate correlation matrices that need to be inverted. In our case, differences between the performances of these two implementations are minor.

Generally, there is no clearly visible residual microseismic peak except for the case of using surface seismometers as witness channels to cancel noise in a 4850\,ft seismometer in Figure~\ref{fig:wiener}B. Thus, we were able to improve over previous results reported in \citet{CoHa2014}. 
Figure~\ref{fig:wiener}A shows that below 0.1\,Hz, the residual almost reaches the limit set by the sensor noise of the Kinemetrics STS-2 broadband seismometers used at the 800\,ft station.
In Figure~\ref{fig:longterm_FIRFFTvsFIR}A, We use equation (\ref{eq:R}) to determine the expected residuals for a few optimal subsets of seismometers taken from the total array. Optimal subsets are the ones that, given a number of seismometers, produce lowest subtraction residuals. The consistency of the expected results with the achieved subtraction indicates the efficacy of our implementation.
The difference between the expected residuals and the true residuals likely relates to a combination of numerical noise when computing the filters, as well as changes in coherence between stations over time.
This also excludes the possibility that the improvement over the previous analysis is simply an increase in the number of channels, as it shows that the expected performance of the Wiener filter rapidly converges as a function of the number of witness sensors, and so it is not simply a gain in signal-to-noise ratio that leads to improved residuals. 

In summary, we can use this method to determine that the underground seismometers significantly increase the accuracy of the measurement of the underground wavefield.
Measurements of this type show the utility of including underground seismometers in future arrays dedicated to time-dependent velocity measurements, allowing predictions at the 1\% level of the wavefield (Figure~\ref{fig:wiener}A), whereas constraining observations to surface stations we are left with at least a 4\% level residual (Figure~\ref{fig:wiener}C).

Figure~\ref{fig:longterm_FIRFFTvsFIR}B shows the efficiency of a Wiener filter calculated one day and applied to data collected on later days. We show the results of a Wiener filter calculated on day 154 that is applied to data later that same day, data the following day, 20 days later, and 37 days later. 
In general, a loss of up to a factor of 2 in the predictive power of the filter can be seen on month-long timescales. 
Some loss in performance is expected, although we note that the subtraction is still better than a factor of 100.
A loss in performance is unsurprising given the changing composition of the seismic field, but the relatively minimal loss in performance indicates that in general, the body-wave vs. fundamental Rayleigh wave content does not have a significant impact on the phase of the correlations measured between the seismometers (which is what determines the composition of the filters). This is because the phase shifts introduced in the seismic time-series predominantly only changes the result if the phase delays introduced are large, which is not the case for an array of this size. This arises from our array dimensions, such that for the wavelengths considered here, the seismometers are well within one wavelength of one another. 
It follows then that the nearby stations are the most important for the subtraction in this case.
Therefore, the difference in phases between body-wave and fundamental Rayleigh waves are not identified.

The Wiener filter can be considered comparable to other coherence or cross-correlation type observations for the time at which it was trained. The filter applied at any other time period should then perform equally well if all aspects of the environment remained constant. When applied to another time period, the fact that there is a difference still between prediction and the actual observation implies that either the intervening medium has changed, or that changes in the ambient noise field cannot be resolved by the array. In our case, we assume that any material velocity changes would be relatively constant over the extent of the different sub-arrays, but still find that different geometries used produce different residuals. The fact that surface-only 2D observations, for example, cannot describe more than 96\% of the waveform (Figure~\ref{fig:wiener}C) implies that there is an upper limit to what we can expect to resolve or explain; that last 4\% may be considered a random level of variability given the geometry used. 

Finally, we can also use the Wiener filter results to test for the presence of low amplitude local sources that have a significant effect on correlations. If this were the case, they would also have a significant effect on our Wiener filters. However, this can be excluded since the Wiener filters prove to be highly efficient with the cancellation of oceanic microseisms (reduction by more than two orders of magnitude in most cases). There are two possibilities for how such excellent subtractions are possible. The first is that the filter is almost fully determined by correlations consistently in phase with microseisms. The other possibility is that a local source produces plane waves consistently in phase with microseisms. Given Homestake's array geometry and the wavelengths of interest, phase differences across the array are small, and therefore local sources are also subtracted to some extent.  However, as the subtraction results correspond to a coherence between target and Wiener filter of about 0.999995, it is very unlikely that a local source produced phase differences that match the ones of the oceanic microseisms so well to achieve the same coherence. It is more likely that local sources were insignificant during the measurements in the relevant frequency range.

\section{Implications for gravitational-wave observations}

This work, and even the deployment of the Homestake array \citep{MaTs2017}, is additionally motivated by open questions in the gravitational-wave community.
Vibrations in the Earth's crust are a significant source of noise in gravitational-wave observatories.
Vibrations in the Earth hinder the precise measurements needed by gravitational-wave detectors, both via the mechanical coupling of vibrations through the mirror supports and via the local gravity fluctuations due to rock density fluctuations, known as Newtonian Noise (NN). While sophisticated seismic-isolation systems are used in order to limit the effect of mechanical couplings \citep{MaEA2015,BrEA2005}, fluctuations in the gravitational field at the interferometer mirrors from local seismic noise and temperature and pressure fluctuations in the atmosphere will be a future limiting noise source below about 20\,Hz \citep{Sau1984,HuTh1998,Cre2008,Har2015}. Wiener filters, combined with knowledge of the wave type, can be used to determine the NN contribution and mitigate its effects.

Understanding the wave content of oceanic microseisms is of high priority for sub-Hz gravitational-wave detectors where seismic fields produce NN about 1000 times stronger than the instrumental noise required to detect gravitational waves \citep{McEA2017}. The measurements of mixed wave type content have significant implications for NN cancellation for potential future low-frequency gravitational-wave detectors. The assumption so far has been that the seismic field is dominated by Rayleigh waves, which greatly helps with the cancellation of the associated NN using off-line Wiener filter subtraction \citep{HaPa2015}. Given that NN cancellation in the presence of multiple wave polarizations is a complicated task even for modest cancellation goals \citep{Har2015}, continuous body-wave content as observed at Homestake would be a substantial additional challenge for plans to suppress seismic NN at sub-Hz frequencies by large factors.
Subtractions at the level of 1\,\% and below do give confidence though that in the case of body-wave and fundamental Rayleigh wave separation, significant mitigation of NN is possible.
We emphasize again that underground seismometers are needed to achieve better than 1\% understanding of the seismic wavefield.
Such capabilities are essential to realize cancellation of terrestrial gravity noise in future gravitational-wave detectors.

\section{Conclusion}
\label{sec:conclusion}

In this paper, we have used one year of data from an underground and surface array deployed in 2015 at the Sanford Underground Research Facility (former Homestake mine) for correlation analyses of the ambient seismic field. The results include the year-long evolution of spectral density and seismometer correlations at 0.2\,Hz and the broadband cancellation of seismic signals in the array using Wiener filters. 
The long-term study of PSDs and correlations at 0.2\,Hz showed evidence of an incessant background of body waves frequently perturbed by week-long Rayleigh-wave transients. These findings are consistent with previous observations, but our findings go beyond previous results as the body-wave content seems to enforce the low-noise model at the Homestake site. This link has not been established before to our knowledge and may apply generally to quiet stations in the continental interior.
Finally, while it has been previously known that array geometry plays an important role in a method's ability to resolve and recover the ambient noise field, our application of Wiener filters allows us to quantify the lower limit of this recovery. These Wiener filters are used to estimate and cancel seismic signals in a target sensor using data from other stations in the array, reducing seismic signals by more than 2 orders of magnitude. By comparing the estimate and residual of different subarrays we find that this can be improved by a factor of 4 by including underground stations to better capture the entire ambient noise wavefield. 

We do note that this characterization of the array geometry and the background ambient noise field may be only one possible application of Wiener filter theory. The exploration of microseismicity remaining after such a prediction and subtraction outlined here may allow the detection of events not possible otherwise. The characterization of site-amplification effects from an array of stations, rather than just a station-station comparison may also be possible. Such topics are beyond the scope of this paper and will be the focus of future work.

Techniques like Wiener filtering, beamforming and observations of coherence decay will continue to be important for quantifying the precision to which seismic velocity measurements can be made, including for observation of temporal changes in seismic velocity structure. These results show that noise correlation studies where Rayleigh waves are usually assumed to be responsible for observations may be contaminated by body waves. Moving forward, the techniques presented here may be useful in larger arrays, and it will be interesting to quantify the degree to which they apply over larger scales.

\acknowledgments
Data used in this project are available from the Incorporated Research Institutions for Seismology (IRIS) \citep{MaTs2014}. 
The seismic instruments used for this array were provided by IRIS through the PASSCAL Instrument Center at New Mexico Tech. 
We thank Nicholas Harmon and an anonymous reviewer for suggestions and improvements to the text.
MC was supported by the David and Ellen Lee Postdoctoral Fellowship at the California Institute of Technology. We thank the staff at the Sanford Underground Research Facility and PASSCAL for assistance, particularly the help of Tom Regan, Jaret Heise, Jamey Tollefson, and Bryce Pietzyk.  This work was supported by National Science Foundation INSPIRE grant PHY1344265. This paper has been assigned LIGO document number LIGO-P1700422.

\bibliographystyle{plainnat}
\bibliography{references}

\end{document}